\documentclass[preprint2]{aastex}

\begin{document}

\title{A search for Fast Radio Bursts at low frequencies with Murchison Widefield Array high time resolution imaging}

\author{S.J. Tingay\altaffilmark{1,2},  C.M. Trott\altaffilmark{1,2}, R.B. Wayth\altaffilmark{1,2}, G. Bernardi\altaffilmark{3,10,18}, J.D. Bowman\altaffilmark{4}, F. Briggs\altaffilmark{5,2}, R.J. Cappallo\altaffilmark{6}, A.A. Deshpande\altaffilmark{7}, L. Feng\altaffilmark{8}, B.M. Gaensler\altaffilmark{9,2,17}, L.J. Greenhill\altaffilmark{10}, P.J. Hancock\altaffilmark{1,2}, B.J. Hazelton\altaffilmark{11}, M. Johnston-Hollitt\altaffilmark{12}, D.L. Kaplan\altaffilmark{13}, C.J. Lonsdale\altaffilmark{6}, S.R. McWhirter\altaffilmark{6}, D.A. Mitchell\altaffilmark{14,2}, M.F. Morales\altaffilmark{11}, E. Morgan\altaffilmark{8}, T. Murphy\altaffilmark{9,2}, D. Oberoi\altaffilmark{15}, T. Prabu\altaffilmark{7}, N. Udaya Shankar\altaffilmark{7}, K.S. Srivani\altaffilmark{7}, R. Subrahmanyan\altaffilmark{7,2}, R.L. Webster\altaffilmark{16,2}, A. Williams\altaffilmark{1}, C.L. Williams\altaffilmark{8}}

\altaffiltext{1}{International Centre for Radio Astronomy Research (ICRAR), Curtin University, Bentley, WA 6102, Australia}
\altaffiltext{2}{ARC Centre of Excellence for All-sky Astrophysics (CAASTRO), Sydney, Australia} 
\altaffiltext{3}{SKA SA, 3rd Floor, The Park, Park Road, Pinelands, 7405, South Africa} 
\altaffiltext{4}{School of Earth and Space Exploration, Arizona State University, Tempe, AZ 85287, USA} 
\altaffiltext{5}{Research School of Astronomy and Astrophysics, Australian National University, Canberra, ACT 2611, Australia} 
\altaffiltext{6}{MIT Haystack Observatory, Westford, MA 01886, USA} 
\altaffiltext{7}{Raman Research Institute, Bangalore 560080, India} 
\altaffiltext{8}{Kavli Institute for Astrophysics and Space Research, Massachusetts Institute of Technology, Cambridge, MA 02139, USA} 
\altaffiltext{9}{Sydney Institute for Astronomy, School of Physics, The University of Sydney, NSW 2006, Australia} 
\altaffiltext{10}{Harvard-Smithsonian Center for Astrophysics, Cambridge, MA 02138, USA}
\altaffiltext{11}{Department of Physics, University of Washington, Seattle, WA 98195, USA} 
\altaffiltext{12}{School of Chemical \& Physical Sciences, Victoria University of Wellington, Wellington 6140, New Zealand} 
\altaffiltext{13}{Department of Physics, University of Wisconsin--Milwaukee, Milwaukee, WI 53201, USA} 
\altaffiltext{14}{CSIRO Astronomy and Space Science (CASS), PO Box 76, Epping, NSW 1710, Australia} 
 \altaffiltext{15}{National Centre for Radio Astrophysics, Tata Institute for Fundamental Research, Pune 411007, India} 
\altaffiltext{16}{School of Physics, The University of Melbourne, Parkville, VIC 3010, Australia} 
\altaffiltext{17}{Dunlap Institute for Astronomy and Astrophysics, University of Toronto, Toronto, ON M5S 3H4, Canada}
\altaffiltext{18}{Department of Physics and Electronics, Rhodes University, PO Box 94, Grahamstown, 6140, South Africa}


\begin{abstract}
We present the results of a pilot study search for Fast Radio Bursts (FRBs) using the Murchison Widefield Array (MWA) at low frequencies (139 - 170 MHz).  We utilised MWA data obtained in a routine imaging mode from observations where the primary target was a field being studied for Epoch of Reionisation detection.  We formed images with 2 second time resolution and 1.28~MHz frequency resolution for 10.5 hours of observations, over 400 square degrees of the sky.  We de-dispersed the dynamic spectrum in each of 372,100 resolution elements of 2$\times$2 arcmin$^{2}$, between dispersion measures of 170 and 675~pc~cm$^{-3}$.  Based on the event rate calculations in \citet{tro13b}, which assumes a standard candle luminosity of $8\times10^{37}$ Js$^{-1}$, we predict that with this choice of observational parameters, the MWA should detect ($\sim10$,$\sim2$,$\sim0$) FRBs with spectral indices corresponding to ($-$2, $-$1, 0), based on a 7$\sigma$ detection threshold.  We find no FRB candidates above this threshold from our search, placing an event rate limit of $<700$ above 700 Jy.ms per day per sky and providing evidence against spectral indices $\alpha<-1.2$ ($S\propto\nu^{\alpha}$).  We compare our event rate and spectral index limits with others from the literature.  We briefly discuss these limits in light of recent suggestions that supergiant pulses from young neutron stars could explain FRBs.  We find that such supergiant pulses would have to have much flatter spectra between 150 and 1400 MHz than have been observed from Crab giant pulses to be consistent with the FRB spectral index limit we derive. 
\end{abstract} 


\section{INTRODUCTION} 
Fast Radio Bursts (FRBs) are enigmatic bursts of radio emission that have high dispersion measures (DM=375 $-$ 1103 pc cm$^{-3}$), $\sim$millisecond durations, $\sim$jansky flux densities, and appear to be localised on the sky at the angular resolution (tens of arcminutes) of the large single dish telescopes with which they have been detected \citep{lor07,kea11,tho13,bur14,pet15,spi14} at frequencies near 1.4 GHz.  These observed properties suggest a cosmological origin, with the large DMs due to the ionised media of the host galaxy, the Milky Way, and the Intergalactic Medium (IGM).  Given this interpretation, rapid energy releases from events involving compact objects are typically invoked as FRB progenitors \citep{sha15,ful14,iwa14,zha14,fal13,han01,uso00,li98,ree77}.  Super-pulses from young pulsars at non-cosmological distances \citep{con15} or cosmological distances \citep{kat15} have also recently been suggested as explanations for FRBs.  The observed frequency dependence of the dispersion and pulse scattering (where seen), and the fluence and DM distributions support a cosmological origin for the bursts \citep{kat15}, although galactic progenitors are also proposed, with the galactic vs extragalactic scenarios vigorously debated \citep{mao15}.  Experiments targeting specific candidate progenitors for FRBs have been conducted in a limited form, for example by \citet{pal14} and \citet{ban12}.

All known FRBs have been detected using large single dish radio telescopes at 1.4 GHz, either Parkes or Arecibo, with all known characteristics of the FRB population determined from these data.  \citet{bur14} examined the Galactic position dependence of known FRBs to assess the evidence for an extragalactic origin.  \citet{pet15} reported the first FRB from real-time data (all previous FRBs were detected in archival databases), allowing an extensive, if ultimately unsuccessful, multi-wavelength follow-up for afterglow emission.  Substantial uncertainties on the basic parameters of the FRB population, such as luminosity distribution, spectral index distribution, and event rate remain \citep{kea15,ran15}.

If unambiguously shown to be cosmological in origin, FRBs would provide an outstanding probe of the conditions of the IGM.  However, in order to take this step, FRBs will have to be localised on the sky with much higher angular resolution than has been possible with single dish radio telescopes.  Angular resolution high enough to identify the FRB with a host galaxy is required, so that redshifts and distances can be obtained.

Interferometric radio telescopes provide much higher angular resolution than single dishes (at a given frequency) and are the natural choice to undertake searches for FRBs with the goal of accurate localisation.  However, the level of complexity introduced by using an interferometer for FRB searches is significant.  The data rates required to search at millisecond timescales are very large, image formation and calibration routines are complex and computationally intensive, and the hardware systems of existing interferometers are rarely optimised for such programs.  In general, there is a trade-off in the processing between coherent searches (maintaining the high data rate, sensitivity, and angular resolution at the expense of computational complexity) and incoherent searches (less sensitive, lower data rates, and worse angular resolution, but greatly reduced computing complexity).  Different programs use these two approaches in complementary ways.  Several such programs, that vary in their use of interferometers for FRB searches, are ongoing.

The V-FASTR project \citep{way11,tho11,way12,tro13a,bur15} uses the Very Long Baseline Array (VLBA) in a commensal mode (between 1.4 and 43 GHz), performing a millisecond timescale incoherent addition of data from all ten antennas to mitigate against radio frequency interference.  The data for any candidate FRBs are reprocessed in an imaging mode for higher sensitivity and milliarcsecond-scale localisation.  To date, no FRBs have been detected by V-FASTR \citep{tro13a} and recent results \citep{bur15} place limits on FRB spectral indices of $\alpha^{\rm 20cm}_{\rm 90 cm}>-6.4$ and $\alpha^{\rm 4 cm}_{\rm 20 cm}<4.0$.  A comparison between V-FASTR and Parkes results provides a limit on the flux density counts, $N(>S)\propto S^{\gamma}$, of $\gamma<-0.5$ with 95\% confidence.

An experiment similar to the V-FASTR experiment, with the large-N, shorter baseline Allen Telescope Array (ATA) covered 150 square degrees of sky over 450 hours at 1.4 GHz and did not detect any FRBs with a sensitivity of 44 Jy in 10 ms \citep{sie12}, corresponding to a fluence of 440 Jy.ms.

An interferometric search for FRBs is being undertaken using the Very Large Array (VLA) at millisecond time resolution at 1.4 GHz \citep{law15}.  In a 166 hour experiment, searching $0~\rm{pc~cm^{-3}} < DM < 3000~\rm{pc~cm^{-3}}$, and with a sensitivity of 120 mJy/beam, no FRBs were detected, corresponding to a fluence limit of 1.2 Jy.ms and an event rate upper limit of $7\times10^{4}$ per sky per day above that fluence.

FRB searches and localisation experiments have generally been undertaken at centimeter and shorter wavelengths.  However, a new generation of highly capable low frequency interferometric radio telescopes have been constructed and commissioned in recent years, notably LOFAR \citep{van14} and the Murchison Widefield Array \citep{tin13,lon09}.  Both of these instruments have established high time resolution capabilities in order to study pulsars and search for FRBs at low radio frequencies \citep{coe14,tre15a,ord14}.  As FRBs have only been detected at relatively high frequencies to date, no empirical information for FRBs at low frequencies has been available until recently.  However, predictions show that the low frequency regime is potentially of great interest for the detection and localisation of FRBs.  Both LOFAR and the MWA have very wide fields of view (primary beam widths up to tens of degrees) and sub-arcminute localisation capabilities.  

\citet{tro13b} predict FRB detection rates with the MWA based on various spectral index and scattering assumptions and find that multiple detections per week of observing are not unreasonable.  \citet{has13} predict FRB detection rates for a range of next-generation radio telescopes, likewise predicting high possible detection rates.  Some models predict that FRBs occur only in a relatively narrow frequency range near 1.4 GHz \citep{iwa14} or that observed FRB rates at 1.4 GHz may be enhanced by extrinsic factors such as interstellar diffractive scintillation (as functions of observing frequency and position on the sky) \citep{mac15}.  Thus, FRB searches over the widest possible frequency range can help understand models of FRB generation and/or propagation.

The first results of FRB searches at low frequencies are starting to become available.  The results of LOFAR pilot surveys using high time resolution capabilities, undertaken using both incoherent methods and coherent methods, have placed the first low frequency FRB event rate limits \citep{coe14}.  The incoherent LOFAR survey (LOFAR Pilot Pulsar Survey: LPPS) places a limit on the FRB event rate of $<$150 day$^{-1}$sky$^{-1}$ for S$>$107 Jy in 1 ms at 142 MHz, corresponding to a fluence of 107 Jy.ms fluence.

Using the UK station of LOFAR and the ARTEMIS processing real-time software backend, \cite{kar15} performed a 1446 hour search for FRBs at DMs up to 320~pc~cm$^{-3}$, finding none and placing an event rate limit of $<29$~sky$^{-1}$~day$^{-1}$ for 5~ms pulses with $S>62$~Jy (fluence of 310 Jy.ms).  From these results, \cite{kar15} derive a limit of $\alpha^{>}_{\sim}+0.5$, assuming power law spectra ($S\propto \nu^{\alpha}$).

In this paper, we describe initial results of an MWA imaging pilot study search for FRBs.  We used the MWA in its routine imaging mode, whereby visibilities with 2 second temporal resolution and 1.28 MHz frequency resolution were used to generate images of the full MWA field-of-view at arcminute angular resolution.  We de-dispersed the dynamic spectrum produced from each resolution element in our images and searched for highly dispersed pulses.  A DM of $500~\rm{cm~pc^{-3}}$ at a centre frequency of 155 MHz with a bandwidth of 30 MHz produces a signal delay across the band of approximately 40 seconds.  Thus, the expected FRB signal temporal sweep is large compared to our time resolution.

The duration of an FRB is expected to be substantially less than our time resolution (given observed scattering at 1.4 GHz and extrapolating to 150 MHz, although there is a range in the scattering properties of FRBs at 1.4 GHz), leading to signal to noise loss.  Such a search is therefore not optimal in terms of its temporal resolution.  However, taking into account imaging sensitivity, field of view, plausible FRB properties \citep{tro13b}, paucity of searches at low frequencies, and potential for arcminute localisation, such an experiment with the MWA is worthwhile.

Using the MWA in its routine imaging mode, we are sensitive to FRBs of approximately 700 Jy.ms at 150 MHz.  The calculations of \cite{tro13b}, applied to the observational parameters relevant to the MWA routine imaging mode (e.g. 2 second integrations), predict that the MWA may be able to detect between 0 and 10 FRBs per 10 hours of observation, depending on the spectral index of the FRB emission.  These expectations are detailed in the next section.  Following the description of the estimated event rates, we describe the observations and data processing for the experiment, the results of the FRB search, a comparison of our results to our expectations, and comparisons between our experiment and other FRB detection experiments.  The results of other FRB searches with the MWA will be reported elsewhere, in particular using the MWA high time resolution voltage capture capabilities \citep{tre15b} and via image plane searches over timescales of tens of seconds \citep{row15}.

\section{EXPECTED EVENT RATES}

The imaging mode for FRB searches was investigated in \citet{tro13b} for the MWA, based on the methodology developed in \citet{tro13a}. The treatment considers the full frequency-dependent primary beamshape and frequency-dependent system noise (sky-dominated at these frequencies); it also includes the loss of sensitivity due to considering broad frequency channels (1.28~MHz), where de-dispersion can only be coarsely performed.  However, with 2 second temporal averaging, frequency-averaging loss is a minimal effect. In addition, in this work we compute and apply the loss of sensitivity due to the $2\times2$ arcmin$^{2}$ image pixels used in imaging, by considering the instrument sampling function (baseline distribution) and average detectability for a source not coincident with the pixel centre, including accounting for the changing angular resolution as a function of frequency across the MWA bandwidth.  

The predictions are based on a standard candle (luminosity of $8\times10^{37}$ Js$^{-1}$) model derived from \citet{lor13}, and calibrated at 1.4~GHz for the more recent Parkes detections reported by \citet{tho13}. \cite{kat15} suggests that the known FRB population and their fluences are consistent with a fixed luminosity population in a Euclidian Universe, supporting the assumptions we have used.  Further details on our model can be found in \citet{tro13b}. Here we consider three values for the \textit{a priori} unknown FRB spectral indices to convert flux densities from 1.4~GHz to the MWA observation frequency of 155~MHz, $\alpha=-2,-1,0$ ($S\propto\nu^{\alpha}$).  This model suggests that the usefully searchable DM range extends up to approximately 700 $\rm{pc~cm^{-3}}$ (Figure 1).

The observations described below were spread over five pointing directions, spanning zenith angles of 0 degrees (zenith) to 38 degrees. The reduced sensitivity of non-zenith pointings (due to geometrical projection of the collecting area) was accounted for in the expected detections, reducing the overall number by a factor of 0.89 compared with pure zenith pointings.  

For the $7\sigma$ detection threshold determined from the data analysis and the observational setup used (both described in the next section), we expect to detect: 0.2 FRBs for a spectral index of $\alpha=0$; 1.6 FRBs for a spectral index of $\alpha=-1$; and 9.3 FRBs for a spectral index of $\alpha=-2$.  

\begin{figure}[ht]
\centering
\includegraphics[width=8cm,angle=0]{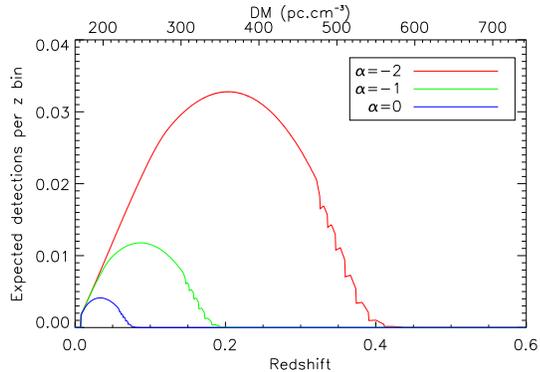}
\caption{Predicted distribution of FRB detections in redshift for each spectral index and a 10.5 hour observation.  The structure in the curves at the high redshift/DM end is due to the sensitivity contribution of the sidelobes of the MWA tile beam, as described in \citet{tro13b}.}\label{fig:predictions}
\end{figure}

Figure \ref{fig:predictions} shows the expected redshift distribution of detections (per $\Delta z=0.0007$) for a 10.5 hour observation and 7$\sigma$ detection threshold, for each spectral index considered.

\section{OBSERVATIONS}
Observations for this experiment were undertaken with the MWA as part of project G0005 (``Search for Variable and Transient Sources in the EOR Fields with the MWA")\footnote{http://mwatelescope.org/astronomers}.  The observations are of one of the fields being studied for the MWA Epoch of Reionisation experiment.  This field, centred at RA=$00$ h; DEC=$-27^{\circ}$ (J2000), is also the subject of searches for radio transients and variables (on timescales of 10s of seconds to years).  The large amount of data accumulated for this field, along with the fact that it has been chosen to be away from the Galactic plane and as free as possible from nearby very bright radio sources, means that it is an ideal field in which to conduct an FRB survey.  In Galactic coordinates, the field centre lies at $l=30$; $b=-78$.  The MWA field of view covers approximately 600 square degrees at 155 MHz.

The observations described here were all obtained in the routine MWA imaging mode, with 2 second correlation integrations and 40 kHz frequency resolution, with dual polarisation correlation.  The 30.72 MHz bandwidth was recorded as 24 $\times$ 1.28 MHz sub-bands, between 138.89 and 169.61 MHz.  Approximately 10.5 hours of data were obtained over four nights of observing on: 2014 October 18 (12:04:40 - 15:24:32 UTC); 2014 November 06 (12:46:24 - 14:09:44 UTC); 2014 November 07 (11:00:48 - 13:53:36 UTC); 2014 November 08 (11:02:56 - 14:01:52 UTC).  Data were collected in a series of 112 second files for each observation period.  The total visibility data volume processed was approximately 5.8 terabytes.

\section{DATA ANALYSIS AND RESULTS}
The raw visibility data were pre-processed through the `cotter' pipeline \citep{2015PASA...32....8O} which performs automated flagging of any radio-frequency interference (RFI) using the `AO Flagger' \citep{2012MNRAS.422..563O} software and converts the data to the standard uvfits format. AO Flagger detects RFI on single MWA baselines and removes features corresponding to thousands of Jy (see, for example, Figure 8 of \citet{2015PASA...32....8O}).  AO Flagger will not, therefore, flag candidate FRBs at the detection levels we are probing in this study.  Subsequent processing was performed with the \textsc{Miriad} \citep{1995ASPC...77..433S} data processing package (Version 1.5, recompiled in order to allow 128 antenna arrays).

\subsection{Imaging and calibration}
An observation of the bright and compact (at MWA angular resolution) radio source 3C 444 using the same instrumental setup was used to calibrate the target field data (assuming a flux density of 79.6 Jy and a spectral index of $-$0.88 at a reference frequency of 160 MHz).  For each target field observation (112 seconds in duration) during a given night, the following process was followed.

The \textsc{Miriad} task \textsc{MFCAL} was used to provide an initial calibration solution from the 3C 444 observation from the same night.  This calibration was transferred to the target field data and applied using \textsc{Miriad} task \textsc{UVAVER}, at the same time averaging in frequency from the native correlated frequency resolution of 40 kHz to 1.28 MHz (the width of an MWA coarse channel).

These calibrated target data were then used to form an image of the target field for each 112 second observation, utilising a single polarisation (XX polarisation).  Only a single polarisation is utilised due to the complexities introduced in the processing by the polarisation dependent primary beam shapes of aperture arrays.  First the data for each 112 second observation were inverted using the \textsc{Miriad} task \textsc{INVERT} (7000x7000 pixel images with 30 arcsec pixel size, robust weighting set to zero, and utilising the multi-frequency synthesis capabilities of \textsc{INVERT}, as well as producing the spectral dirty beam with a beam four times the area of the image produced).  The \textsc{Miriad} task \textsc{MFCLEAN} was then used to clean the image produced by \textsc{INVERT}, using 50,000 iterations and only cleaning the inner 66\% of the image.  \textsc{MFCLEAN} produced a model of the radio sources in the field that was then used to self-calibrate the target field visibilities using \textsc{Miriad} task \textsc{SELFCAL} (solving for amplitude and phase components of the complex gain for each 1.28 MHz sub-band in the data).  Figure 2 shows an example of a resulting image.

\begin{figure*}[ht]
\centering
\includegraphics[width=13cm,angle=270]{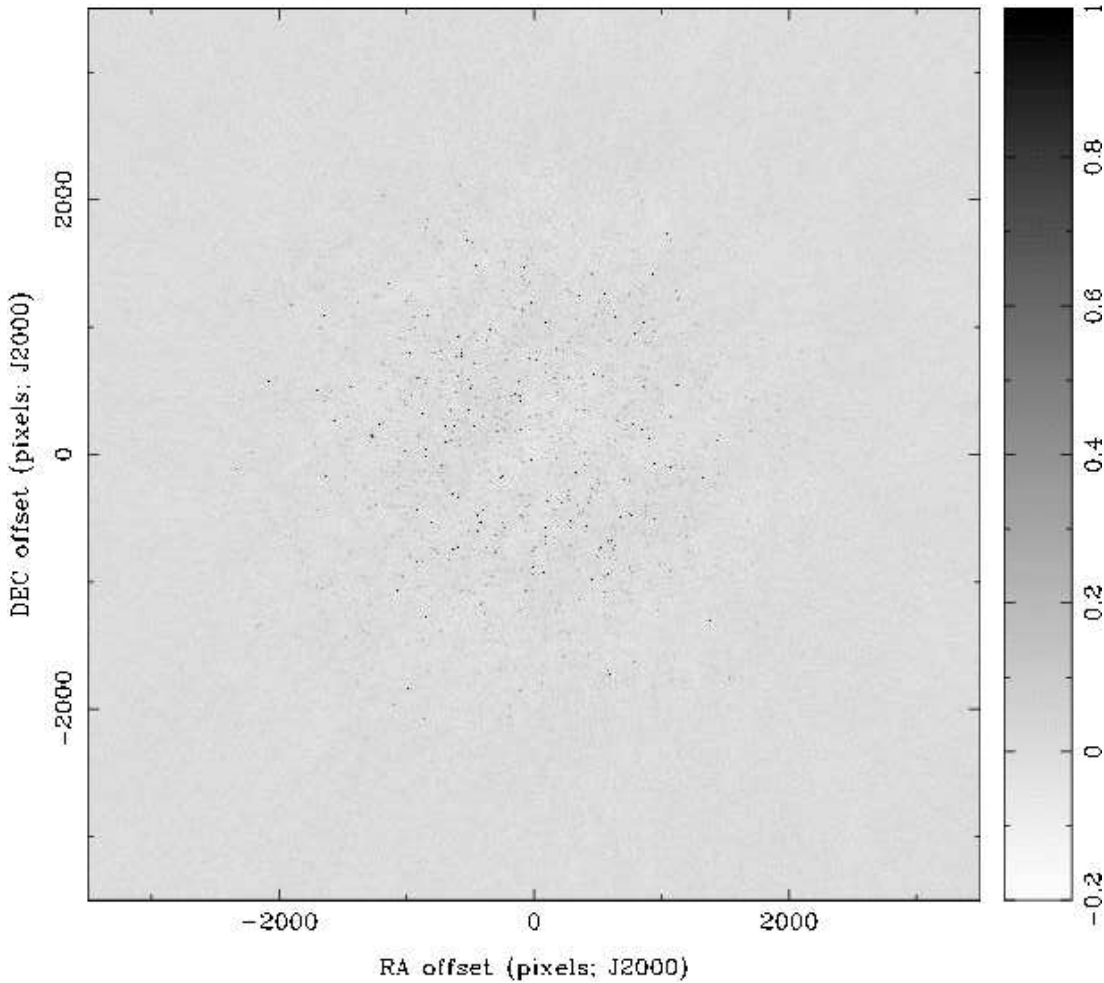}
\caption{Example image of target field following imaging and calibration procedure described in the text.  The image has a restoring beam of $221\arcsec \times 194\arcsec$ at PA $-74^{\circ}$.  The pixels in the image are $0.5~{\rm arcmin} \times~0.5~{\rm arcmin}$ in extent, giving the image a $58^{\circ} \times 58^{\circ}$ field of view.  The greyscale bar on the right is in units of Jy/beam.}
\end{figure*}

\subsection{Formation of dynamic spectra}
The \textsc{Miriad} task \textsc{UVMODEL} was used to subtract the model of the field derived from \textsc{MFCLEAN} from the self-calibrated visibilities for each 112 second dataset.  This step was taken to subtract bright and persistent radio sources from the data before performing high resolution imaging and the search for FRBs.

For FRBs at our 7$\sigma$ detection limit of $\sim$15 Jy.ms we are in no danger of removing them from our data via the subtraction of the target field model.  In an image produced from a 112 second dataset, a 15 Jy.ms FRB would appear as an approximate 40 mJy persistent source.  The level to which the target field model is cleaned is approximately 120 mJy, a factor of three greater than the level at which a 15 Jy.ms FRB would appear.  FRBs with detections of significance up to $\sim20\sigma$ in our pipeline will not be affected by the subtraction of the target field model.  Above this level of significance, the significance of detections will be underestimated but detections will not be eliminated.  

The model-subtracted visibility data were imaged for each 2 second time step and for each 1.28 MHz frequency sub-band, in the form of a spectral cube.  For each 2 second time step, \textsc{INVERT} was once again used (with $750\times750$ pixels and 120 arcsec pixel sizes), producing images at each of the $24\times1.28$ MHz frequency sub-bands.  Due to the method by which \textsc{Miriad} generates image cubes (retaining a fixed representation of the point spread function for all frequencies, but scaling the dirty images at each frequency), it was necessary to regrid the image cubes to a fixed image size and pixel size, such that a fixed pixel is located at a fixed celestial coordinate at all frequencies.  Due to this effect, only the inner $610\times610$ pixels of the images are usable for the FRB search, as the edge pixels are progressively lost moving from the low frequency edge of the band to the high frequency edge.  An example image cube, representing a single 2 second time step and all 24 frequency channels is shown in Figure 3.

\begin{figure*}[ht]
\centering
\includegraphics[width=13cm,angle=270]{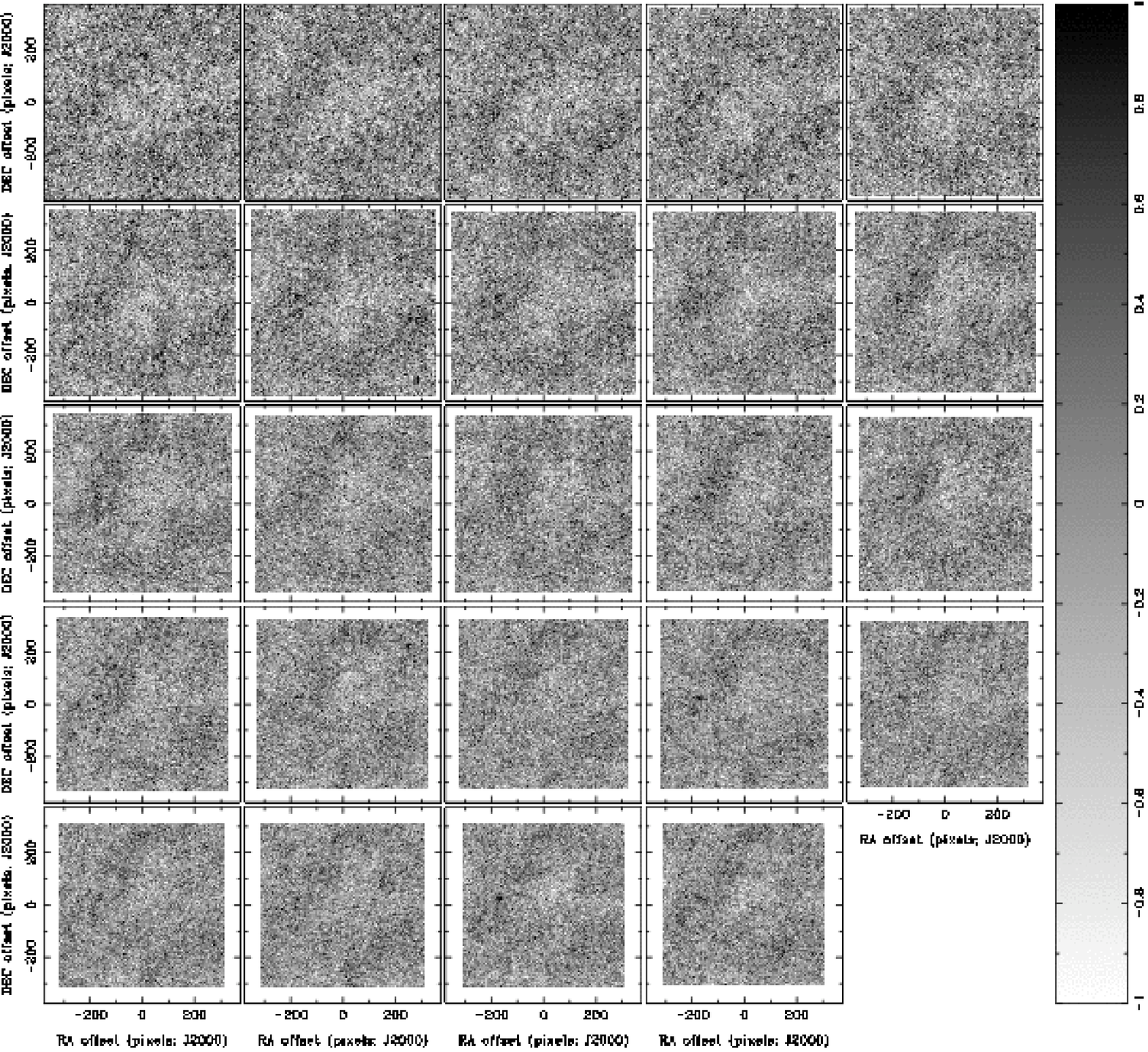}
\caption{Example images of target field as a function of frequency following subtraction of the model derived from Figure 2, for a single 2 second integration, as described in the text.  The lowest frequency image appears at top left and images increase in frequency (in steps of 1.28 MHz) from left to right and down the page to the highest frequency image at bottom right.  The greyscale bar on the right is in units of Jy/beam.}
\end{figure*}

The image root mean square (RMS) for the 24 example images shown in Figure 3 range between 280 mJy and 380 mJy.  This is broadly in line with expectations.  \citet{tin13} quote a theoretical 1$\sigma$ point source sensitivity for full bandwidth (30.72 MHz) observations, two polarisations, and 26 second integrations of 10 mJy.  The images in Figure 3 have 1/24 of the full bandwidth, a single polarisation, and 2 second integrations.  Thus, the expected image RMS would be approximately 270 mJy, within a factor of order unity of the values measured from Figure 3.  Uncleaned diffuse Galactic emission may contribute to the measured image RMS.  In the final de-dispersed time series, we would expect the RMS in the time series to be a factor of $\sqrt{24}$ lower, which is what is realised (see below).

The \textsc{Miriad} task \textsc{IMLIST} was used to extract the pixel values from the image cubes as a function of frequency for each 2 second time step.  Thus, after processing one observation file of 112 second duration, dynamic spectra for $610\times610$ pixels were generated with 2 second resolution (56 time steps) and 1.28 MHz frequency resolution (24 frequency steps over 30.72 MHz bandwidth) for a single polarisation.  

These dynamic spectra were then aggregated in time over the course of the full observation period, filling missing time steps and any gaps between 112 second observation files with zeros in each frequency sub-band.  A typical dynamic spectrum for one pixel would then contain  approximately $24\times3600=86,400$ measurements per 2 hours.  Over the full field of view ($610\times610$ pixels), the aggregate number of measurements in the dynamic spectra would then typically be $24\times3600\times610\times610=3.2\times10^{10}$.  An example dynamic spectrum for one pixel is shown in Figure 4.  These dynamic spectra are the input for the de-dispersion and event detection analysis.  Typically some very small proportion of data (small handful of 2 second time steps) are corrupted for each night of observation, evidenced by spurious values in all image pixels.  These time steps were identified and flagged by replacing the values in the dynamic spectra with zeros for those time steps.

\begin{figure}[ht]
\centering
\includegraphics[width=5cm,angle=270]{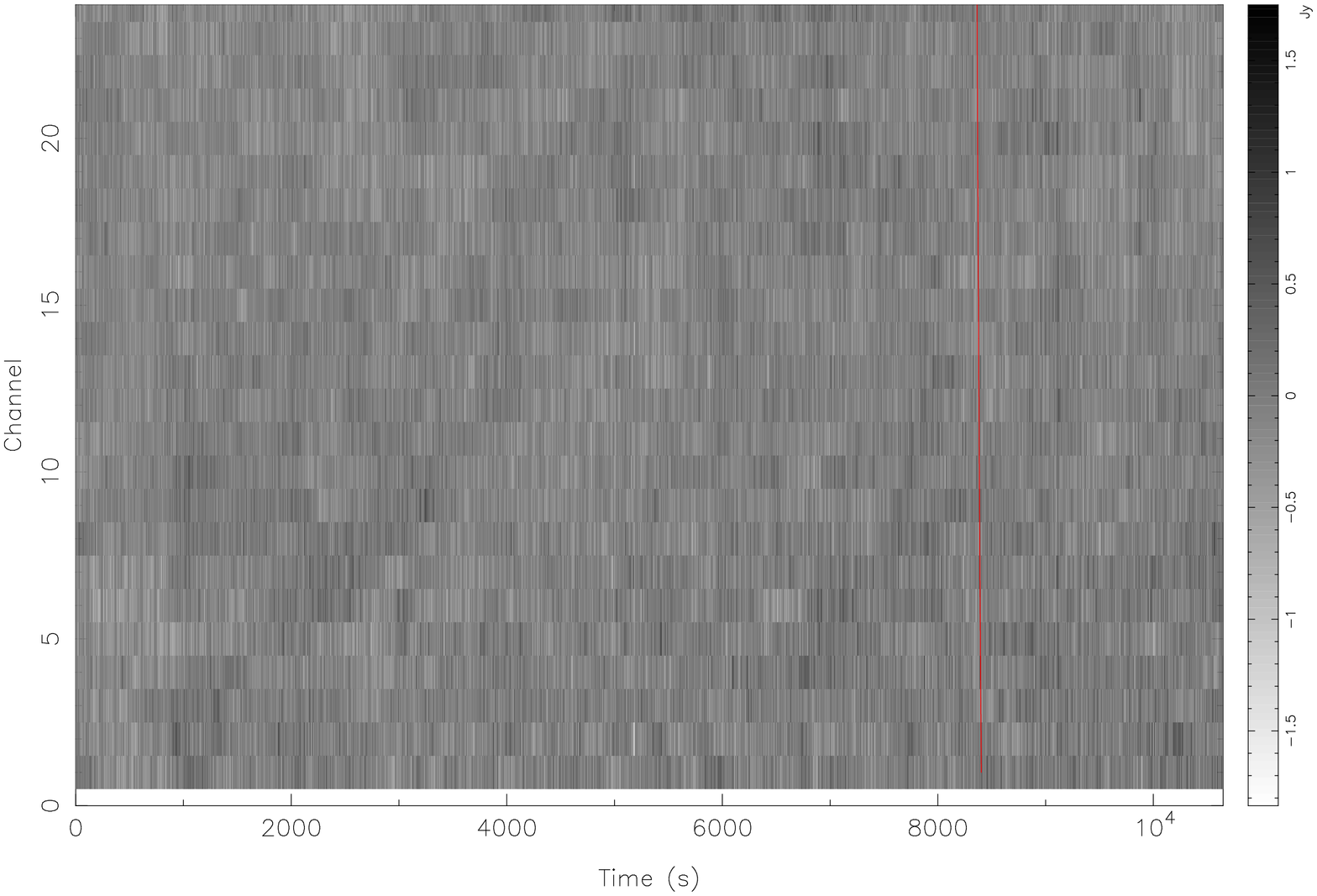}
\includegraphics[width=5cm,angle=270]{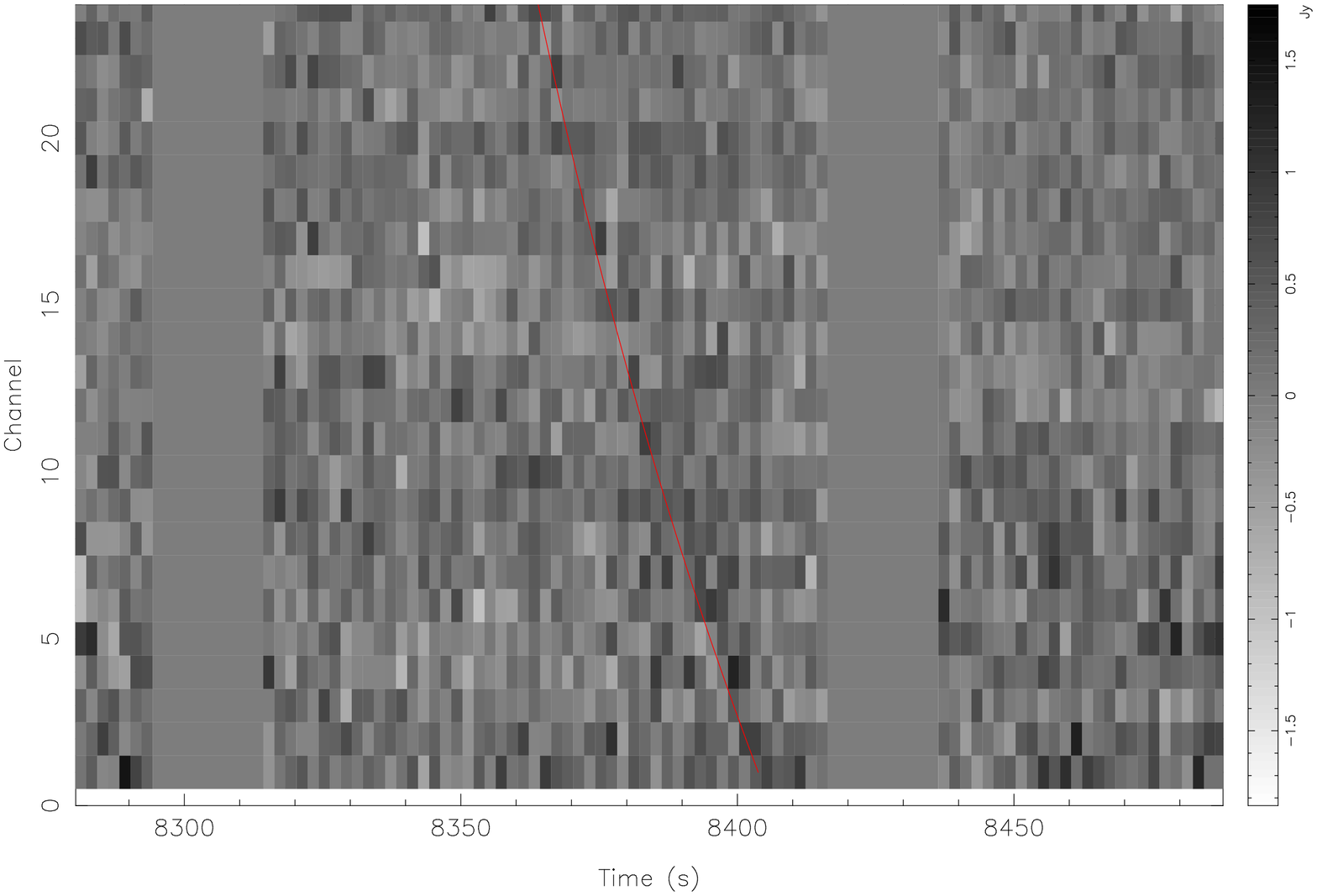}
\caption{An example dynamic spectrum for pixel (513,138) from the observations of 2014 November 07, showing a $6\sigma$ candidate detection 8364 s from the start of the observation, detected at DM=594.08 $\rm{pc~cm^{-3}}$.  The top panel shows the full time range.  The bottom panel shows the time range zoomed in around the candidate event.  Gaps between files are evident as zeroes.  The red lines represent the dispersion sweep across the band for the candidate event DM.  The greyscale bar on the right is in units of Jy/beam.}
\end{figure}

\subsection{De-dispersion and FRB search}
The dynamic spectrum for each pixel over the course of each night of observation was then de-dispersed and the de-dispersed time series were searched for single pulse events.  The range of DMs (170 $-$ 675 pc cm$^{-3}$) was determined at the high end by the values suggested by the modelling of \cite{tro13b}, illustrated in Figure 1, and set at the low end to avoid galactic objects and contamination by interplanetary scintillation of strong, compact radio emission from active galactic nuclei.

It is worth noting that the predictions of \cite{tro13b}, and hence the DM range searched, are model-dependant but based on sensible assumptions.  However, some other plausible models (e.g., that FRBs are not standard candles or that measured DM is not from IGM) would have a different rate constraints due to missed FRBs at high DM or different scattering.

The de-dispersion was performed using DART \citep{way11}.  DART was configured to produce de-dispersed time series over the DM list: 170.00; 201.62; 233.87; 266.77; 300.32; 334.55; 369.46; 405.07; 441.39; 478.44; 516.23; 554.77; 594.08; 634.19; and 675.09.  These DMs were chosen for a bandwidth spanning 138.89 - 169.61 MHz and used the same progression scheme as \cite{pal14}, where it was shown that the chance of triggering adjacent DMs with noise fluctuations is small.  Thus, for each pixel de-dispersed, 20 de-dispersed time series were generated, one for each DM trial.

Using equations 3 and 4 from \citet{cor03} to estimate the temporal smearing due to de-dispersion error, we find that at the bottom edge of our DM range, the maximum temporal smearing experienced is $<$0.2 seconds.  That is, less than 10\% of our integration time in the worst case and less at other DMs.  Thus, the sensitivity losses due to the choice of DM progression are insignificant.

Each de-dispersed time series was then searched for single pulse events.  The subtraction of the model of the field described above is not perfect, resulting in small variations in the power in each pixel.  A perfect subtraction would result in dynamic spectra with zero mean Gaussian noise, plus short timescale transient events (e.g. FRBs).  In turn, the de-dispersed time series would also consist of zero mean Gaussian noise.  In practise, small variations in the quality of the model subtraction with time shift the mean value of the dynamic spectra (and likewise the de-dispersed time series) away from zero, although the noise remains very close to Gaussian.  This effect is accounted for in the search of the de-dispersed time series by calculating the mean over 100 time steps (200 seconds) and calculating the RMS around that mean to assess the significance of any time sample's deviation from the mean value.  The typical timescale for variations in the mean power is $\sim$500 seconds.

A threshold of $5\sigma$ was used to record candidate single pulse events from the de-dispersed time series, in order to assess the noise statistics.  A two hour observation at 2 second resolution, across $610\times610$ pixels, and with 20 de-dispersed time series, amounts to a notional set of $2.7\times10^{10}$ independent trails.  Notionally, given independent trials and Gaussian statistics, one would expect approximately 7,500 events with power more than $5\sigma$ higher than the mean, 25 events more than $6\sigma$ higher than the mean, and no events more than $7\sigma$ above the mean.  Table 1 shows the number of events at $5\sigma$ and $6\sigma$ detected from our different datasets, according to the above processing scheme, as well as the expected number of events assuming fully independent trials and Gaussian statistics.

From Table 1, one can see that the number of events recorded from the data are consistent with what is expected from notional independent trials and Gaussian noise.  A set of competing and complex effects in the data affect the noise statistics.   We outline some of these effects below.

First, the effect noted above due to variations in the mean power levels will have residual effects over the 200 second timescales used to evaluate the mean and RMS in the de-dispersed time series, meaning that the RMS will be marginally higher than simply due to Gaussian noise. Detections at a reported significance level will therefore correspond to a slightly higher significance level in practise.  This effect slightly changes the number of recorded events, relative to the number of expected events based on Gaussian statistics.  

Second, it has been previously noted as a result of the work done here, described in \cite{kap15}, that this search for FRBs is also sensitive to the detection of the effects of interplanetary scintillation (IPS), the apparent short timescale variation of radio source flux density as the result of radio wave propagation through the solar wind.  IPS causes low level variations on 2 second timescales and may contribute to the RMS in the de-dispersed time series for different pixels in the field at different times.  Occasionally these variations can be extreme and can trigger very high significance detections in the FRB search pipeline at multiple low value DMs \citep{kap15}.  

Third, residual radio frequency interference may affect the RMS of the de-dispersed time series, although this has been shown to be at very low levels for the MWA \citep{2015PASA...32....8O}.  However, we do flag a very small fraction of our data in the dynamic spectra here and some RFI below the easily identifiable levels may persist. 

Finally, the mean noise level is estimated from only 100 data samples, due to the fluctuations in power on longer timescales, leading to sample variance uncertainty on the expected number of false positives. This effect is quantified in order to provide the errors on the false positive rates in Table 1.

Given these complex effects may be present in our data, the agreement between the nominal expectation and the results of the FRB pipeline in Table 2 should be considered reasonable.  We visually inspect all candidate events reported above $6\sigma$ and can verify that in all cases listed in Table 2, the $6\sigma$ detections are consistent with an apparent random alignment of noise fluctuations triggering the relevant DM.  An example of such a case is shown in Figure 4 (dynamic spectrum) and Figure 5 (corresponding de-dispersed time series).  The de-dispersed time series in Figure 5 has units of intensity (Jy/beam), as the data are derived from pixel-based images.  We assume point source emission from FRBs, in which case hereafter we use units of flux density (Jy).

Confidence in the overall performance of the FRB search pipeline was established through the blind detection of IPS in the dataset of 2014 November 06, where 100\% amplitude variations on 2 second timescales in the apparent flux density of PKS 2322$-$275 triggered $>10\sigma$ detections in the low DM channels of the search of the de-dispersed time series (these detections of IPS have been discarded from the analysis above).  The IPS detection via the pipeline verifies that the construction of pixel-based dynamic spectra from the image cubes has been correctly handled and that the detection stage of the pipeline functions correctly.  The detailed description of the IPS detection is fully described elsewhere \citep{kap15}.  Further, examples such as seen in Figure 5 clearly illustrate that the de-dispersion step is picking up and accumulating positive noise fluctuations that align with the correct dispersion sweep across the band.

\begin{figure}[ht]
\centering
\includegraphics[width=5cm,angle=270]{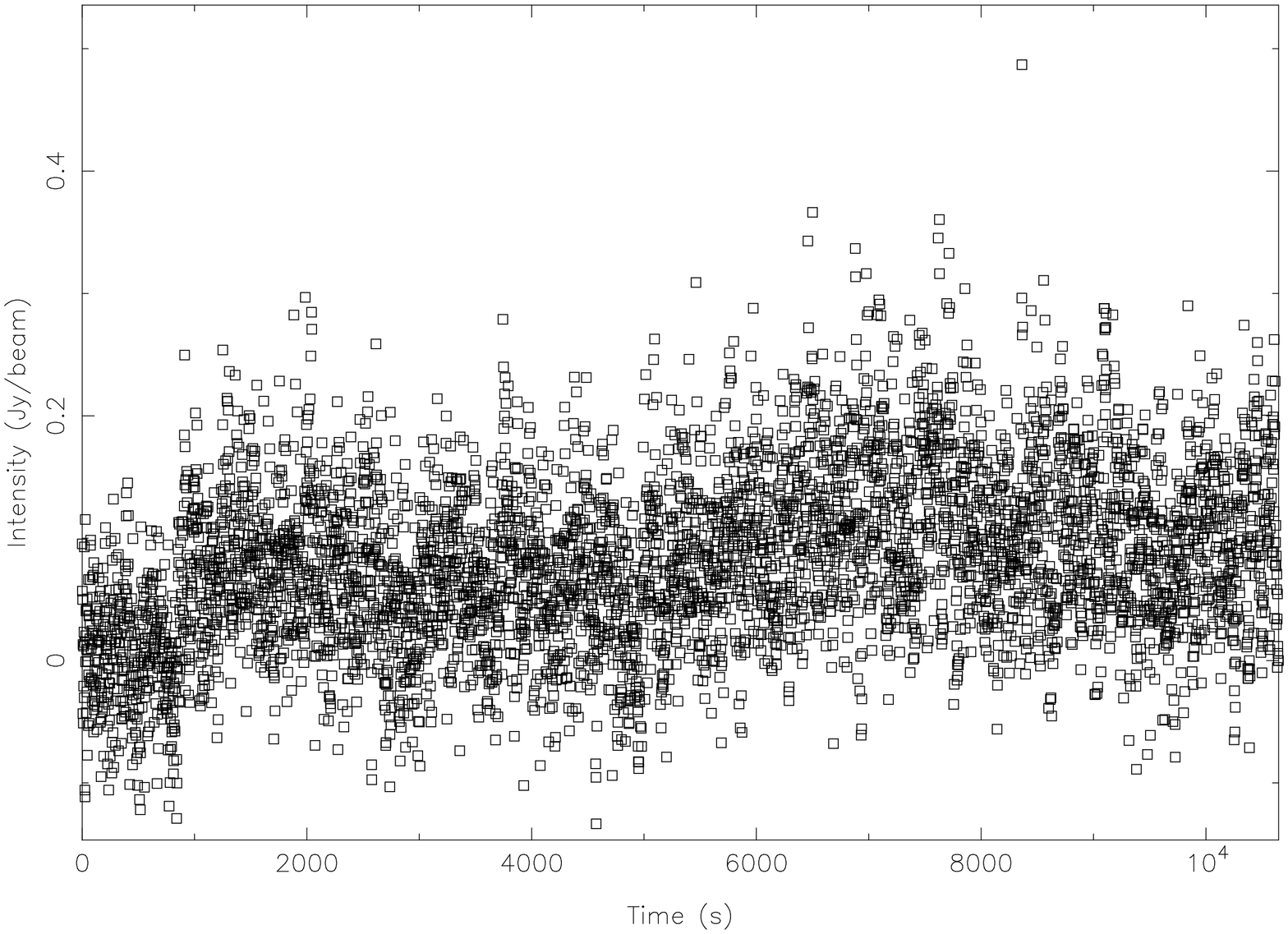}
\caption{An example de-dispersed time series for pixel (513,138) (from data in Figure 4) in which a $6\sigma$ event was detected at DM=594.08 $\rm{pc~cm^{-3}}$, 8364 s from the start of the observation.}
\end{figure}

\begin{table*}[ht]
  \begin{tabular}{c c c c c c c} \hline
    \#   Date & Observation  & Number of trials      & Expected events & 5$\sigma$ events & Expected events & 6$\sigma$ events  \\
              & duration &                       & above 5$\sigma$ & detected         & above 6$\sigma$ & detected \\ \hline
         2014/10/18&03:20:32&$\sim4.5\times10^{10}$&$\sim13\pm^{15}_{13}\times10^{3}$&7,422&$44\pm^{74}_{44}$&17 \\
         2014/11/06&01:23:20&$\sim1.9\times10^{10}$&$\sim5\pm^{6}_{5}\times10^{3}$&5,026&$18\pm^{31}_{18}$&24 \\
         2014/11/07&02:52:48&$\sim4.0\times10^{10}$&$\sim12\pm^{13}_{12}\times10^{3}$&7,150&$38\pm^{63}_{38}$&20 \\
         2014/11/08&02:58:56&$\sim4.0\times10^{10}$&$\sim12\pm^{13}_{12}\times10^{3}$&8,454&$38\pm^{63}_{38}$&56 \\ \hline
 \end{tabular}
   \caption{Expected number of 5σ and 6σ events in the case of independent trials and Gaussian noise (for a fixed, known noise level), against the number of 5σ and 6σ events detected by the pipeline. Errors (1σ) on expected numbers incorporate both noise-level uncertainties in the data and sample variance. No 7σ events are expected or detected.}
\end{table*}

\section{DISCUSSION AND CONCLUSION}

Given that our rates of detection at various levels of significance match the expectations from Gaussian statistics (Table 1), and that our visual inspection of all $6\sigma$ events does not reveal any compelling signals, we conclude that the highest significance events we detect, all less than $7\sigma$, are consistent with noise and adopt this as the simplest explanation of our detections.  Thus, we place a detection limit of $7\sigma$ for single pulse events in our de-dispersed time series that defines our basic result $-$ no FRB detections above $7\sigma$.  The typical RMS around the mean in our de-dispersed time series is approximately 50 mJy, giving a $7\sigma$ detection limit in 2 seconds of approximately 350 mJy, corresponding to a $7\sigma$ limit on the fluence of 700 Jy.ms. Table \ref{table:results} reports the significance of each null detection. The $p$-value describes the probability of detecting zero FRBs when $k$ are expected, where the expected values are taken from Section 2, and the significance describes the confidence with which we can reject the null hypothesis (computed according to a Poisson rate of FRB events in time and position). Given these significance values, a spectral index of $\alpha=-2$ is rejected at high significance ($>$99\%), and $\alpha=-1$ is rejected at moderate confidence (79\%).

\begin{table}[ht]
\centering
  \begin{tabular}{r | c c} \hline
    $\alpha$ & $p$ & Significance \\\hline
    $-2$ & 9$\times$10$^{-5}$ & $>$99\% \\
    $-1$ & 0.21 & ~79\% \\
    $0$ & 0.81 & ~19\%\\   \hline
 \end{tabular}
   \caption{$p$-value of detecting zero events when $k$ are expected, given by $\exp{(-k)}$, and significance of rejecting the null hypothesis (rejecting a model with a given spectral index).}\label{table:results}
\end{table}

The MWA FRB results can be compared with other recent reports of null detections at low frequencies.  \citet{coe14} describe the first results from the LOFAR LPPS search for FRBs at 150~MHz. Despite having more hours on sky, their smaller field of view and poorer sensitivity yield a comparable experiment to that performed here. 

Scaling our experiment's field of view to the $4\pi$ Steradians of the full sky over a 24 day, we place a 2$\sigma$ upper bound on the rate of FRBs of ($\alpha=0$),

\begin{equation}
{\rm R_{FRB}} \left(S>700 \,\,{\rm Jy.ms}\right) < 700 \,\, {\rm /sky/day}.
\end{equation}

Thus, at 95\% confidence ($2\sigma$), we can reject the hypothesis that we should have observed three FRBs, when zero were actually observed. This corresponds to a spectral index limit of $\alpha{>}-1.2$, assuming the \cite{tro13b} model (based on FRBs as standard candles) we use for prediction.

We have compared our limit to other previous results, assuming a spectral index of $\alpha=0$, in order to compare our limit with those obtained at higher frequencies.  We consider the recent results of the VLA FRB search by \citet{law15}, the recent low frequency ARTEMIS limit from \citet{kar15}, and results from V-FASTR \citep{tro13a}.  The limit we find here is consistent with previous limits, in particular within the interpretation of a homogenous, stationary population of objects represented by $N(>S)\propto S^{-3/2}$, as described by \citet{coe14}.

\cite{con15} propose an extragalactic but non-cosmological origin for FRBs, that they are supergiantpulses from pulsars formed from recent core collapse supernovae in galaxies within a few hundred Mpc, with the large DMs explained by the dense and ionised medium of the young supernova remnants.  This idea could be testable by preferentially targeting FRB searches towards galaxies in which recent core collapse supernovae have been discovered.  From the list of all known supernovae maintained by the Central Bureau for Astronomical Telegrams (CBAT\footnote{https://www.cbat.eps.harvard.edu/lists/Supernovae.html}), eight core collapse supernovae have been discovered in galaxies within the MWA field observed here, since 1991.  Thus, the extremely wide field of view of the MWA would be advantageous in tests of the suggestions of \cite{con15}, since it is capable of targeting significant numbers of known and recent supernovae simultaneously.

In terms of the physical model put forward by \cite{con15}, as well as that proposed by \cite{kat15}, based on supergiantpulses from young core collapse supernovae, the spectral index limits we and others have recently derived are interesting.  In this work, we place a 95\% confidence limit of $\alpha>-1.2$ between the Parkes frequency and the MWA frequency, using the underlying assumptions of \cite{tro13b}.  \cite{kar15} place a limit of $\alpha^{>}_{\sim}+0.5$, but only for relatively low DMs.  The broadest band study to date, by \cite{bur15}, places 95\% limits of $\alpha^{\rm 20cm}_{\rm 90 cm}>-6.4$ and $\alpha^{\rm 4 cm}_{\rm 20 cm}<4.0$.

Giant pulses have been detected from a small number of pulsars.  The best studied giant pulses are from the Crab pulsar which is approximately 1000 years old.  Giant pulses have also been observed from B0540$-$69, J1939$+$2134, and some millisecond and long period pulsars (see references in \citet{oro15}).  For the Crab, \citet{oro15} performed simultaneous observations of giant pulses with the MWA at 193 MHz and Parkes at 1382 MHz; they find 23 giant pulses simultaneously detected at both frequencies, ranging in spectral index between $\alpha=-3.5$ and $\alpha=-4.9$ ($S\propto\nu^{\alpha}$).

We note that the limits on FRB spectral indices from the results presented here, $\alpha>-1.2$ between 150 MHz and 1400 MHz, and the results from \cite{kar15} $\alpha>+0.5$ over the same frequency range, both rule out the observed spectral indices for Crab giant pulses over the same frequency range from \cite{oro15}.  For supergiant pulses from young neutron stars to be a viable explanation for FRBs, the superpulses would have to have much flatter spectra than has been observed for Crab giant pulses in the frequency range 150 - 1400 MHz.

As a pilot study using a novel FRB search approach with the MWA, a useful constraint on the FRB population has been established from only 10.5 hours worth of data.  This work has used existing software packages and non-parallelised computing.  In order to fully utilise this proof-of-concept approach, the processing of approximately two orders of magnitude more data ($\sim$1000 hours) would be very useful.  No FRB detections in this search mode after 1000 hours would place stricter constraints on the event rates under the model assumptions used here, ruling out a spectral index of $\alpha=0$ at $>99\%$ level.  

Given that the proof-of-concept software pipeline described here is relatively inefficient ($\sim$3 days to process $\sim2$ hours of data on a single processing core) and requires some manual intervention, a more efficient processing strategy is required in order to process substantially more data.  For example, \textsc{MIRIAD}, while fit for purpose for this work, does not include features that would further optimise the results, including w-projection.  This work is ongoing, motivated by the potential returns.

The imaging mode for FRB searching is an interesting compromise between incoherent methods and fully coherent methods.  Fully coherent beamforming involves processing massive data rates and forming a very large number of beams to cover the MWA field of view.  This is computationally very demanding, requiring bespoke software solutions.  Incoherent beamforming naturally recovers the full MWA field of view, but at a significantly lower sensitivity and with the loss of localisation.  The imaging approach recovers almost the full coherent sensitivity of the array, recovers the full field of view, preserves localisation potential, and can be implemented (albeit somewhat inefficiently) using existing interferometric software packages.  Thus, the imaging approach is attractive as a relatively simple way to make progress.  These conclusions are also supported by similar work at the VLA \citep{law15}.

Given the uncertainties regarding the characteristics of the FRB population, and the fact that these events have only ever been detected at 1.4 GHz with large single dish radio telescopes, undertaking searches as deeply as possible and over the largest possible fields of view, and at frequencies significantly below and above 1.4 GHz, are a clear imperative.  The Phase 1 low frequency Square Kilometre Array (SKA) will go significantly beyond the MWA or LOFAR in terms of sensitivity and will be located at the same site as the MWA.  If the mysteries of FRBs are not all solved by the time the SKA commences science observations, the SKA will have a lot to contribute.  

The MWA, as the low frequency precursor to the SKA, is in a good position to undertake pathfinder FRB searches at low frequencies from the same site as the eventual SKA, informing FRB search strategies for the SKA.

\section{Acknowledgements}
CMT is supported by an Australian Research Council Discovery Early Career Researcher Fellowship (DE140100316).  This scientific work makes use of the Murchison Radio-astronomy Observatory, operated by CSIRO. We acknowledge the Wajarri Yamatji people as the traditional owners of the Observatory site. Support for the MWA comes from the U.S. National Science Foundation (grants AST-0457585, PHY-0835713, CAREER-0847753, and AST-0908884), the Australian Research Council (LIEF grants LE0775621 and LE0882938), the U.S. Air Force Office of Scientific Research (grant FA9550-0510247), and the Centre for All-sky Astrophysics (an Australian Research Council Centre of Excellence funded by grant CE110001020). Support is also provided by the Smithsonian Astrophysical Observatory, the MIT School of Science, the Raman Research Institute, the Australian National University, and the Victoria University of Wellington (via grant MED-E1799 from the New Zealand Ministry of Economic Development and an IBM Shared University Research Grant). The Australian Federal government provides additional support via the Commonwealth Scientific and Industrial Research Organisation (CSIRO), National Collaborative Research Infrastructure Strategy, Education Investment Fund, and the Australia India Strategic Research Fund, and Astronomy Australia Limited, under contract to Curtin University. We acknowledge the iVEC Petabyte Data Store, the Initiative in Innovative Computing and the CUDA Center for Excellence sponsored by NVIDIA at Harvard University, and the International Centre for Radio Astronomy Research (ICRAR), a Joint Venture of Curtin University and The University of Western Australia, funded by the Western Australian State government. 

{\it Facility:} \facility{MWA}.

\end{document}